\documentclass[12pt,twoside]{article}
\usepackage{fleqn,espcrc1,epsfig}

\def\rsim{\mathrel{\raise2pt\hbox to 8pt{\raise -5pt\hbox{$\sim$}\hss{$>$}}}}
\def\lsim{\mathrel{\raise2pt\hbox to 8pt{\raise -5pt\hbox{$\sim$}\hss{$<$}}}}

\begin {document}

\title{Electro-weak Capture Reactions for Astrophysics}
\author{R.\ Schiavilla
\address{Jefferson Lab, Newport News, Virginia 23606 \\
         $\>\>$and \\
         $\>\>$Department of Physics, Old Dominion University,
         Norfolk, Virginia 23529}}
\maketitle

\begin{abstract}
The status of {\it ab initio} microscopic calculations of
the $^2$H($p,\gamma$)$^3$He and $^3$He($p,e^+\nu_e$)$^4$He
reactions is reviewed.  The methods used to generate
accurate nuclear ground- and scattering-state wave functions, and
to construct realistic electro-weak transition operators are
described.  The uncertainties in the theoretical predictions, particularly
those relevant to the $p\,^3$He weak capture, are
discussed.  For the $d$$p$ radiative capture, the theoretical
results are compared with the TUNL data in the energy range 0--100 keV.
\end{abstract}

\section{Introduction}

In the present talk I will review the progress made in the last couple of
years in {\it ab initio} microscopic calculations of the 
$^2$H($p,\gamma$)$^3$He~\cite{Viv00} and
$^3$He($p,e^+\nu_e$)$^4$He~\cite{Mar00} capture reactions.
These reactions provide a sensitive testing ground for models of
nuclear interactions and currents.

The outline of the talk is as follows.  I will first describe
the correlated-hyperspherical-harmonics method used to obtain accurate wave functions
for the $A$=3 and 4 bound and scattering states.  Next, I will discuss
the model for the nuclear electromagnetic current, and will review our current
understanding of the radiative capture reaction
$^2$H($p$,$\gamma$)$^3$He.  In the third part, I will discuss
the model for the nuclear weak current, and present results for the proton
weak capture on $^3$He.  This process has lately
received considerable attention, due to the 
Super-Kamiokande collaboration measurements of the energy spectrum
of electrons recoiling from scattering with solar neutrinos~\cite{Fuk99}.

The theoretical description of few-nucleon capture reactions
constitutes a challenging problem from the standpoint of nuclear
few-body theory.  For example, since the $A$=3 and 4 bound states
are approximate eigenstates of the magnetic moment operator,
the corresponding transition matrix elements between these
and the initial $d$$p$ (or $d$$n$) and $n\,^3$He states, occurring
in the radiative captures, vanish due to orthogonality.  As a
result, these capture processes are extremely sensitive to: (i) small
components in the wave functions, particularly the D-state admixtures
generated by tensor interactions, (ii) many-body terms in the electro-weak
current operator, and (iii) P-wave capture contributions.  
\section{Wave Functions}

The nuclear Hamiltonian used in the calculations reported
here has the form

\begin{equation}
H= \sum_i \frac{{\bf p}_i^2}{2\,m} + \sum_{i<j} v_{ij}
  +\sum_{i<j<k} V_{ijk} \ ,
\end{equation}
where the two-nucleon interaction is the Argonne $v_{18}$ (AV18)
model~\cite{WSS95}, and the three-nucleon interaction is the
Urbana-IX (UIX) model~\cite{Pud95}.  This Hamiltonian predicts reasonably well
the low-lying energy spectra of systems with $A \leq 8$ nucleons in \lq\lq exact\rq\rq
Green's function Monte Carlo calculations~\cite{Wir00}.  The experimental
binding energies of the trinucleons and $\alpha$ particle are exactly 
reproduced, while those of the $A$=6--8
systems are underpredicted by a few percent. 
This underbinding becomes (relatively) more and more severe as the
neutron-proton asymmetry increases.  An additional failure of the AV18/UIX
model is the underprediction of spin-orbit splittings in the excitation
spectra of these light systems.  These failures 
have in fact led to the development
of new three-nucleon interaction models.  These developments
as well as a discussion of a number of issues regarding
two-nucleon interactions, such as non-localities, unitary equivalence, etc.,
can be found in Friar's~\cite{Fri00} and
Pandharipande's~\cite{Pan00} contributions to these proceedings. 

While the nuclear interaction models above are simple to write
down, bound- and scattering-state solutions for light nuclei have proven
to be rather difficult to obtain.  Intense work in this area and the
continuing increase in computational capabilities have led, by now,
to the development of a number of methods, each with different
strengths and domains of applicability (for a review and an
assessment of them, see Ref.~\cite{Car98}).  The
Faddeev-Yakubovsky and quantum Monte Carlo
methods are reviewed, respectively, by
Gl\"ockle~\cite{Glo00} and Pandharipande~\cite{Pan00}
in these proceedings.  Here, I will briefly discuss
the correlated-hyperspherical-harmonics (CHH) technique,
as implemented by the Pisa group for the $A$=3~\cite{KRV93}
and $A$=4~\cite{VKR95} nuclei, and summarize a number of
results obtained for the bound-state
properties and low-energy scattering parameters.

In essence, the CHH method consists in expanding the wave functions on a
suitable basis, and in determining the expansion coefficients variationally.
As an example, the trinucleon bound-state wave function is written as

\begin{equation}
\Psi=\sum_{\alpha} \frac{u_{\alpha}(\rho)}{\rho^{5/2}}
\sum_{{\rm cyclic} \> ijk} Z_{\alpha} (i;jk) \>\>,
\label{psi3}
\end{equation}
where the hyperradius $\rho=\sqrt{x_i^2+y_i^2}$
(${\bf x}_i$ and ${\bf y}_i$ are the Jacobi variables), and the
known functions $Z_{\alpha}(i;jk)$ are antisymmetric under the
exchange $j \rightleftharpoons k$ and
account for the angle ($\hat{\bf x}_i$ and
$\hat{\bf y}_i$), spin ($s_i$ and $s_{jk}$),
isospin ($t_i$ and $t_{jk}$), and
hyperangle ($\phi_i={\rm cos}^{-1}x_i/\rho$) dependence of
channel $\alpha$.  Correlation factors, which account for
the strong state-dependent correlations
induced by the nucleon-nucleon interaction, are included
in the functions $Z_{\alpha}(i;jk)$.
The Rayleigh-Ritz variational principle,
$\langle \delta_u \Psi | H-E_0 | \Psi\rangle = 0$,
is used to determine the ground-state energy $E_0$
and the functions $u_{\alpha}(\rho)$.  Carrying
out the variations with respect to the $u_{\alpha}$'s
leads to a set of coupled second-order differential equations, which
are then solved by standard numerical techniques.

The $N$$d$ cluster wave function $\Psi^{LSJJ_z}$ (again, as
an example), having incoming orbital angular
momentum $L$ and channel spin $S$ coupled to total $JJ_z$, is expressed as

\begin{equation}
   \Psi^{LSJJ_z}  = \Psi_{C}^{JJ_z}
   + \Psi_{A}^{LSJJ_z} \ ,\label{eq:scatte}
\end{equation}
where the term $\Psi_C$ vanishes in the limit of large
intercluster separation, and hence describes the system
in the region where the particles are close to
each other and their mutual interactions are large.
The term $\Psi_A^{LSJJ_z}$, instead, describes the system
in the asymptotic region, and contains the dependence upon
the $R$-matrix elements, from which phase-shifts and mixing
angles are obtained.  The \lq\lq core\rq\rq\ wave function
$\Psi_C$ is expanded in the same basis as the bound-state wave
function $\Psi$, and both the $R$-matrix elements and functions
$u_{\alpha}(\rho)$ occurring in the expansion of $\Psi_C$ are
determined by making use of the Kohn variational principle.
Results for the $^3$He and $^4$He binding energies, and $p\,^3$He
scattering lengths, obtained with the CHH method, are listed
in Table~\ref{tb:scl}.
\begin{table}[htb]
\caption{Binding energies, $B_3$ and $B_4$, of $^3$He and $^4$He,
and $p\,^3$He singlet and triplet S-wave scattering lengths, $a_{\rm s}$ and
$a_{\rm t}$, calculated with the CHH method
using the AV18 and AV18/UIX Hamiltonian models.}
\label{tb:scl}

\begin{tabular}{@{}lllll}
\hline
Model      & $B_3$(MeV)  & $B_4$(MeV) & $a_{\rm s}$(fm) & $a_{\rm t}$(fm) \\
\hline
AV18       & 6.93 & 24.01 & 12.9   & 10.0   \\
AV18/UIX   & 7.74 & 27.89 & 11.5   &  $\>\>$9.13   \\
\hline
EXP        & 7.72 & 28.3 &10.8$\pm$2.6~\protect\cite{AK93}
                  & $\>\>$8.1$\pm$0.5~\protect\cite{AK93} \\
           &      &        &         & 10.2$\pm$1.5~\protect\cite{TEG83}\\
\hline
\end{tabular}\\[2pt]
\end{table}
\section{The Nuclear Electromagnetic Current}

The nuclear current operator consists of one- and many-body terms
that operate on the nucleon degrees of freedom:

\begin{equation}
{\bf j}({\bf q})= \sum_i {\bf j}^{(1)}_i({\bf q})
             +\sum_{i<j} {\bf j}^{(2)}_{ij}({\bf q})
             +\sum_{i<j<k} {\bf j}^{(3)}_{ijk}({\bf q}) \ ,
\end{equation}
where ${\bf q}$ is the momentum transfer, and
the one-body operator ${\bf j}^{(1)}_i$ has the
standard expression in terms of single-nucleon
convection and magnetization currents.  The two-body current
operator has \lq\lq model-independent\rq\rq and
\lq\lq model-dependent\rq\rq components (for a review, see Ref.~\cite{Car98}).
The model-independent terms are obtained from the charge-independent part
of the AV18, and by construction satisfy current conservation
with this interaction.  The leading operator is the isovector
\lq\lq $\pi$-like\rq\rq current obtained
from the isospin-dependent spin-spin and tensor interactions.
The latter also generate an isovector \lq\lq $\rho$-like \rq\rq current, while
additional model-independent isoscalar and isovector currents arise from the
central and momentum-dependent interactions.  These currents are short-ranged
and numerically far less important than the $\pi$-like current.  Finally,
models for three-body currents have been derived in Ref.~\cite{Mar98}, however
the associated contributions have been found to be very small in studies
of the magnetic structure of the trinucleons~\cite{Mar98}. 

The model-dependent currents are purely transverse
and therefore cannot be directly linked to the underlying
two-nucleon interaction.  Among them, those associated with
the $\Delta$-isobar are the most important ones in the
momentum-transfer regime being discussed here.  These currents are
treated within the transition-correlation-operator
(TCO) scheme~\cite{Mar98,Sch92}, a scaled-down
approach to a full $N$+$\Delta$ coupled-channel treatment.
In the TCO scheme, the $\Delta$ degrees of freedom
are explicitly included in the nuclear wave functions by writing

\begin{equation}
\Psi_{N+\Delta}=\left[{\cal{S}}\prod_{i<j}\left(1\,+\,U^{TR}_{ij}\right)
\right]\, \Psi \ ,
\label{eq:psiNDtco}
\end{equation}
where $\Psi$ is the purely nucleonic component, $\cal{S}$ is the
symmetrizer and the transition correlations $U^{TR}_{ij}$ are
short-range operators, that convert $NN$ pairs
into $N\Delta$ and $\Delta\Delta$ pairs.
In the results reported here, the $\Psi$
is taken from CHH solutions of the AV18/UIX Hamiltonian with nucleons only
interactions, while the $U^{TR}_{ij}$ is obtained from two-body
bound and low-energy scattering state solutions of the full $N$-$\Delta$
coupled-channel problem.  Both $\gamma N \Delta$ and $\gamma \Delta \Delta$
$M_1$ couplings are considered with their values,
$\mu_{\gamma N \Delta}=3$ n.m.\ and
$\mu_{\gamma \Delta \Delta}=4.35$ n.m., obtained from data~\cite{Sch92}.
\section{The $p$$d$ Radiative Capture}

There are now available many high-quality data, including differential
cross sections, vector and tensor analyzing powers, and photon polarization
coefficients, on the $pd$ radiative capture at c.m.\ energies
ranging from 0 to 2 MeV~\cite{Sea96,Mea97,Wea99,SK99}.  These data indicate
that the reaction proceeds predominantly through S- and P-wave capture.
The aim here is to verify the extent to which they can be described
satisfactorily by a calculation based on a realistic Hamiltonian
(the AV18/UIX model) and a current operator constructed consistently
with the two- and three-nucleon interactions~\cite{Viv00}.
\begin{figure}[bth]
\centerline{
\epsfig{file=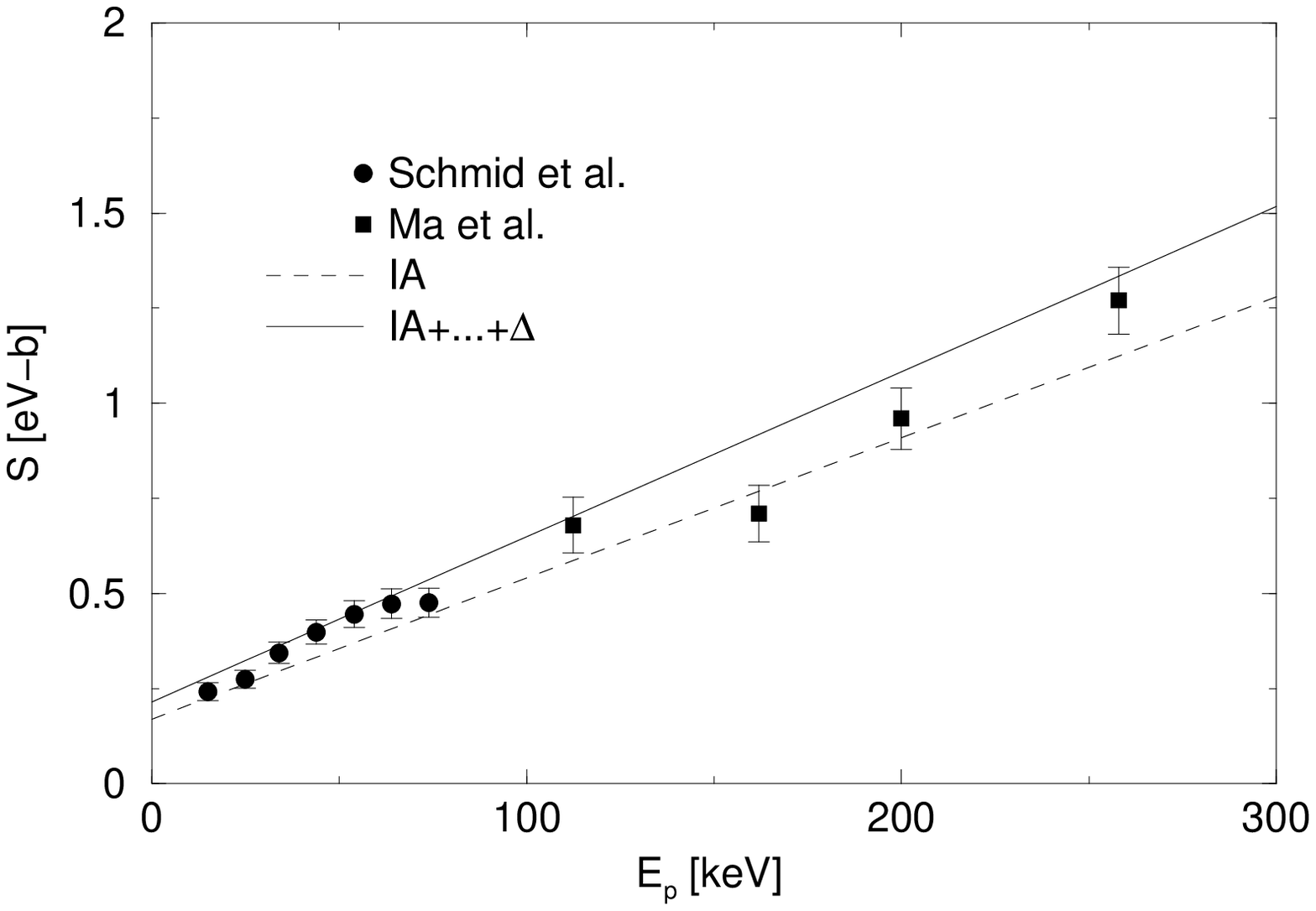,height=3in,width=3in}
\epsfig{file=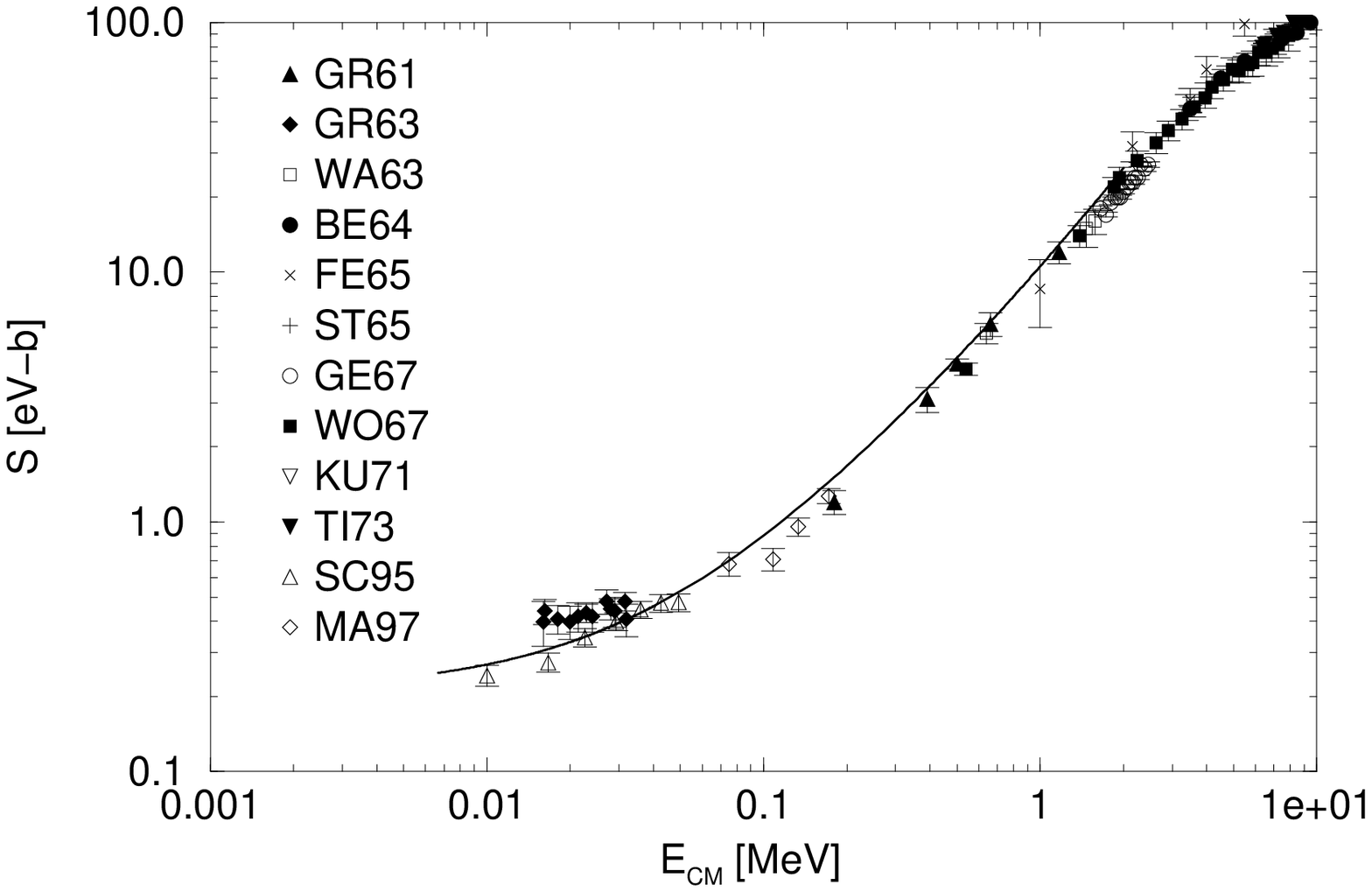,height=3in,width=3in}}
\caption{The $S$-factor for the ${}^2$H($p$,$\gamma$)${}^3$He reaction
in the c.m. energy range 0--2 MeV,
obtained with the AV18/UIX Hamiltonian model and one-body only (dashed line)
or both one- and many-body (solid line) currents.  In the right panel, only the
results of the full calculation are shown.}
\label{fig:S}
\end{figure}
\begin{figure}[bth]
\centerline{
\epsfig{file=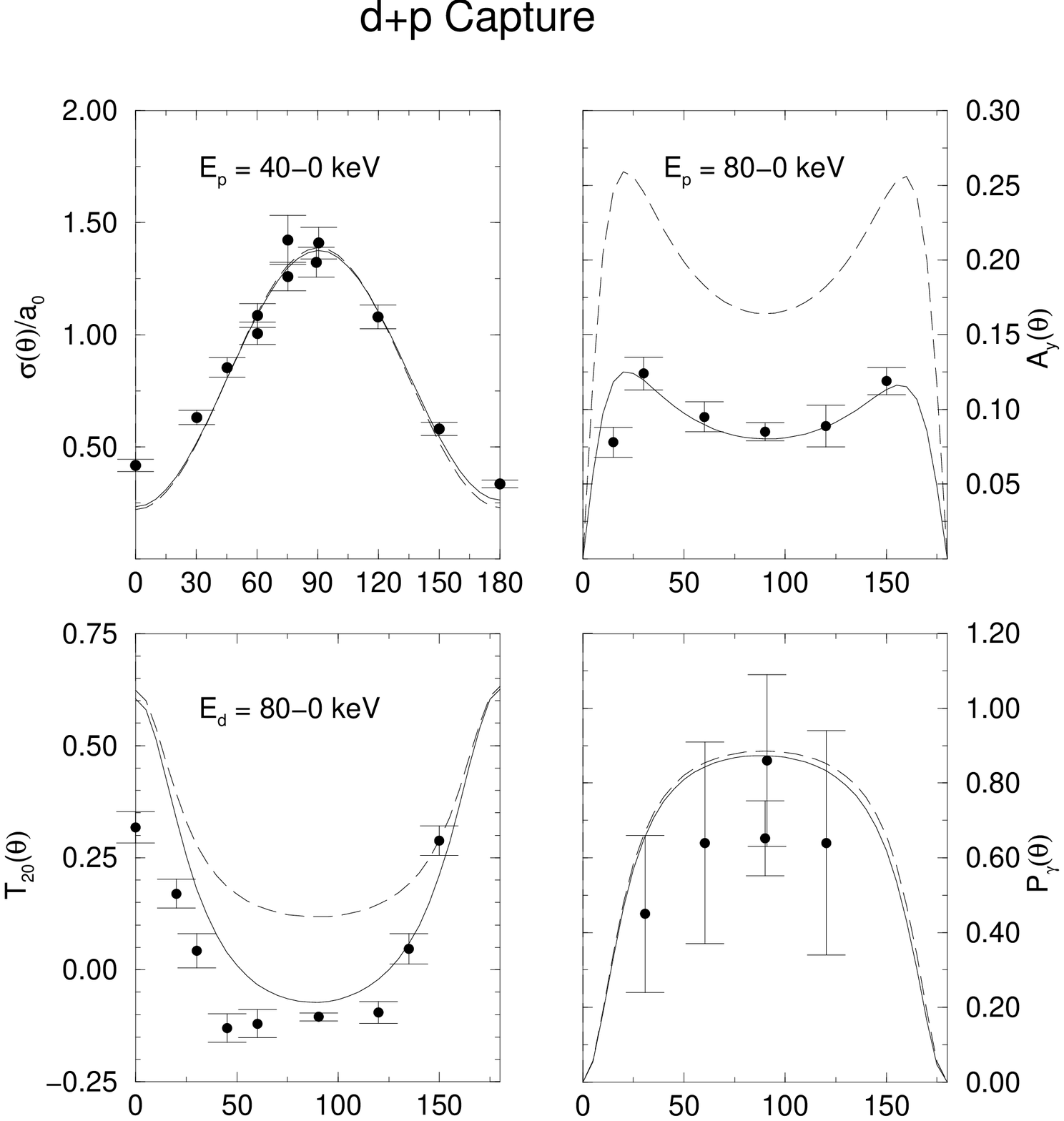,height=3.5in,width=5in}}
\caption{The energy integrated cross section $\sigma(\theta)/a_0$
($4\pi a_0$ is the total cross section), vector analyzing power
$A_y(\theta)$, tensor analyzing power $T_{20}(\theta)$ and photon
linear polarization coefficient $P_\gamma(\theta)$ obtained with the
AV18/UIX Hamiltonian model and one-body only (dashed line)
or both one- and many-body (solid line) currents
are compared with the experimental results of Ref.~\protect\cite{Sea96}.}
\label{fig:cpt080keV}
\end{figure}

The calculated $S$-factor in the c.m.\ energy
range 0--2 MeV is compared with data in Fig.~\ref{fig:S},
while the predicted angular distributions of the differential
cross section $\sigma(\theta)$, vector and tensor analyzing powers
$A_y(\theta)$ and $T_{20}(\theta)$, and photon
linear polarization coefficient $P_\gamma(\theta)$ are compared with
the TUNL data below 50 keV from
Refs.~\cite{Sea96,Wea99} in Fig~\ref{fig:cpt080keV}.
The agreement between the full theory, including many-body
current contributions, and experiment is generally good. 
However, a closer inspection of the figures reveals the
presence of significative discrepancies between theory and experiment
in the $S$-factor below 40 keV, and in the small angle behavior of
$\sigma(\theta)$ and $T_{20}(\theta)$.

The S-wave capture proceeds mostly through the $M_1$ transitions connecting
the doublet and quartet $pd$ states to $^3$He--the associated reduced
matrix elements (RMEs) are denoted by $m_2$ and $m_4$, respectively.
The situation for P-wave capture is more complex, although at energies
below 50 keV it is dominated by the $E_1$ transitions
from the doublet and quartet $pd$ states having channel spin
$S$=1/2, whose RMEs we denote as $p_2$ and $p_4$.  The $E_1$
transitions involving the channel spin $S=3/2$ states, while
smaller, do play an important role in $T_{20}(\theta)$.

The TUNL~\cite{Wea99} and Wisconsin~\cite{SK99} groups have determined
the leading $M_1$ and $E_1$ RMEs via fits to the measured observables.
The results of this fitting procedure are compared with the
calculated RMEs in Table~\ref{tab:rmew}.
The phase of each RME is simply related to
the elastic $pd$ phase shift~\cite{SK99}, which at these low energies
is essentially the Coulomb phase shift.
As can be seen from Table~\ref{tab:rmew},
the most significant differences between theoretical and experimental RMEs
are found for $|p_4|$.  The theoretical overprediction of $p_4$
is the cause of the discrepancies mentioned above in the
low-energy ($\le 50$ keV) $S$-factor and small
angle $\sigma(\theta)$. 

It is interesting to analyze the ratio
$r_{E1} \equiv |p_4/p_2|^2$.
Theory gives $r_{E1}\simeq 1$, while from
the fit it results that $r_{E1}\approx 0.74\pm0.04$.
It is important to stress that the calculation of these RMEs is not
influenced by uncertainties in the two-body currents, since their
values are entirely given by the long-wavelength form of the $E_1$
operator (Siegert's theorem), which has no spin-dependence (for a thorough
discussion of the validity of the long-wavelength approximation
in $E_1$ transitions, particularly suppressed
ones, see Ref.~\cite{Viv00}).  It is therefore
of interest to examine more closely the origin of the above discrepancy.
If the interactions between the $p$ and $d$ clusters are switched off,
the relation $r_{E1} \simeq 1$ then simply follows
from angular momentum algebra.
Deviations of this ratio from one are therefore
to be ascribed to differences induced by the interactions in the
$S$=1/2 doublet and quartet wave functions.
The AV18/UIX interactions in these channels do not change
the ratio above significantly.  It should be emphasized that
the studies carried out up until now ignore, in the continuum states, the
effects arising from electromagnetic interactions
beyond the static Coulomb interaction between protons.
It is not clear whether the inclusion of these
long-range interactions, in particular their spin-orbit component,
could explain the splitting between
the $p_2$ and $p_4$ RMEs observed at very low energy.
Note that this discrepancy seems to disappear at 2 MeV~\cite{Viv00}.
\begin{table}[htb]
\caption{Magnitudes of the leading $M_1$ and $E_1$ RMEs
for $pd$ capture at $E_p=40$ keV.}
\label{tab:rmew}

\begin{tabular}{@{}llll}
RME & IA & FULL & FIT \\
\hline
 $|m_2|$ & 0.172  & 0.322 & 0.340$\pm$0.010  \\
 $|m_4|$ & 0.174  & 0.157 & 0.157$\pm$0.007  \\
 $|p_2|$ & 0.346  & 0.371 & 0.363$\pm$0.014  \\
 $|p_4|$ & 0.343  & 0.378 & 0.312$\pm$0.009  \\
\hline
\end{tabular}\\[2pt]
\end{table}

Finally, the doublet $m_2$ RME is underpredicted by theory at the
5 \% level.  On the other hand, the cross section for $n$$d$
capture at thermal neutron energy is calculated to be 578 $\mu$b
with the AV18/UIX model, which is 15 \% larger than the experimental
value (508$\pm$15) $\mu$b~\cite{JBB82}.
Of course, $M_1$ transitions, particularly
doublet ones, are significantly influenced by many-body current
contributions.  Indeed, an analysis of the
isoscalar ($\mu_S$) and isovector ($\mu_V$)
magnetic moments of the trinucleons~\cite{Mar98}
suggests that the present model
for the isoscalar two-body currents, constructed from the AV18 spin-orbit
and quadratic-momentum dependent interactions, tends to overestimate
$\mu_S$ by about 5 \%.  The experimental value for $\mu_V$, however, is almost
perfectly reproduced.  The present model for two-body isoscalar
currents needs to be improved.  
\section{The Nuclear Weak Current}

The nuclear weak current and charge operators
consist of vector and axial-vector parts, with
corresponding one- and many-body components.  The weak
vector current and charge
are constructed from the corresponding (isovector)
electromagnetic terms, in accordance with the
conserved-vector-current hypothesis, and thus
have~\cite{Mar00} \lq\lq model-independent\rq\rq
and \lq\lq model-dependent\rq\rq components.  The former 
are determined by the interactions, the latter
include the transverse currents associated with $\Delta$ excitation.

The leading many-body terms in the
axial current, in contrast to the case of the
weak vector (or electromagnetic) current, are those due to $\Delta$
excitation, which are treated within the TCO scheme, discussed above.
The axial charge operator includes
the long-range pion-exchange term~\cite{Kub78}, required by low-energy
theorems and the partially-conserved-axial-current relation,
as well as the (expected) leading short-range terms constructed
from the central and spin-orbit components of the nucleon-nucleon
interaction, following a prescription due to Riska and
collaborators~\cite{Kir92}.

The largest model dependence is in the weak axial current.  
The $N$$\Delta$ axial coupling constant
$g_{A}^{*}$ is not well known.  In the quark-model, it is
related to the axial coupling constant
of the nucleon by the relations $g_{A}^{*}=(6\sqrt{2}/5) g_A$.
This value has often been used in the
literature in the calculation of $\Delta$-induced axial current contributions
to weak transitions.  However, given the uncertainties inherent
to quark-model predictions, a more reliable estimate for $g_A^*$
is obtained by determining its value phenomenologically.
It is well established by now~\cite{Sch98}
that one-body axial current
lead to a $\simeq$ 4 \% underprediction of the measured
Gamow-Teller matrix element in tritium $\beta$-decay.
This small 4 \% discrepancy can then be used
to determine $g_{A}^{*}$~\cite{Mar00}.  While this
procedure is inherently model dependent, its actual
model dependence is in fact very weak, as has been
shown in Ref.~\cite{Sch98}.
\section{The $p\,^3{\rm He}$ Weak Capture}

The $hep$ capture process is induced by the weak interaction
Hamiltonian $H_W$.  After partial-wave expansion of the
$p\,^3$He scattering state, the transition amplitude is
written as~\cite{Mar00}

\begin{equation}
\langle \,^4{\rm He}|H_{W}|p\,^3{\rm He};s_1,s_3 \rangle=G_V
\sum_{LSJJ_z} C_{s_1 s_3 J_z}^{LSJ} \,
\langle\Psi_{4}| l^\sigma j_\sigma^{\dag}({\bf q})|
\Psi_{1+3}^{LSJJ_z}\rangle  \ , \label{eq:fhwi2}
\end{equation}
where $l^\sigma$ and $j_\sigma({\bf q})$ are the lepton
and nuclear weak currents, respectively, and $C_{s_1 s_3 J_z}^{LSJ}$
denotes products of Clebsch-Gordan coefficients.  The study
reported in Ref.~\cite{Mar00} includes all transitions
connecting the $p\,^3$He S- and P-wave channels to the $^4$He bound state.
The corresponding wave functions are obtained from realistic
Hamiltonians consisting of the AV18/UIX and older AV14/UVIII~\cite{WSA84}
interaction models.  

The calculated values for the astrophysical $S$-factor
in the energy range 0--10 keV
are listed in Table~\ref{tb:sfact}.  Inspection of the
table shows that: (i) the energy dependence is
rather weak, the value at $10$ keV is only about 4 \% larger
than that at $0$ keV; (ii) the P-wave capture states are found to
be important, contributing about 40 \% of the calculated
$S$-factor.  However, the contributions from D-wave channels
are expected to be very small, as explicitly
verified in $^3$D$_1$ capture.
(iii) The many-body axial currents
associated with $\Delta$ excitation play a crucial
role in the (dominant) $^3$S$_1$ capture, where they reduce
the $S$-factor by more than a factor of four; thus
the destructive interference between the one- and many-body
current contributions, obtained in Ref.~\cite{Sch92}, is
confirmed in the study of Ref.~\cite{Mar00}, based on more accurate
wave functions.  The (suppressed) one-body contribution
comes mostly from transitions involving the D-state
components of the $^3$He and $^4$He wave functions, while
the many-body contributions are predominantly due to transitions
connecting the S-state in $^3$He to the D-state in $^4$He, or viceversa.
\begin{table}[htb]
\caption{The $hep$ $S$-factor, in units of $10^{-20}$ keV~b, calculated
with CHH wave functions corresponding to the AV18/UIX Hamiltonian model,
at $p\,^3$He c.m.\ energies $E$=0, 5, and 10 keV.  The rows
labelled \lq\lq one-body\rq\rq and \lq\lq full\rq\rq list the
contributions obtained by retaining the one-body only and both
one- and many-body terms in the nuclear weak current.  The contributions due
the $^3$S$_1$ channel only and all S- and P-wave channels are
listed separately.}
\label{tb:sfact}

\begin{tabular}{@{}lllllll}
\hline
& \multicolumn{2}{c} {$E$=$0$ keV} &
  \multicolumn{2}{c} {$E$=$5$ keV} &
  \multicolumn{2}{c} {$E$=$10$ keV} \\
\hline
& $^3$S$_1$ & S+P & $^3$S$_1$ & S+P & $^3$S$_1$ & S+P\\
\hline
one-body  &26.4  & 29.0 & 25.9 & 28.7 & 26.2 & 29.3 \\
full      &6.38  & 9.64 & 6.20 & 9.70 & 6.36 & 10.1 \\
\hline
\end{tabular}
\end{table}

The chief conclusion of Ref.~\cite{Mar00} is
that the $hep$ $S$-factor is predicted to be $\simeq$ 4.5
times larger than the value adopted in the
standard solar model (SSM)~\cite{BBP98}.  This enhancement,
while very significant, is smaller than that first suggested
in Ref.~\cite{BK98}.  Even though this result is
inherently model dependent, it is unlikely that the model dependence
is large enough to accommodate a drastic increase in the value obtained here.
Indeed, calculations using Hamiltonians based on the AV18 two-nucleon
interaction only and the older AV14/UVIII two- and
three-nucleon interactions predict zero energy $S$-factor values of
$12.1 \times 10^{-20}$ keV~b and $10.2 \times 10^{-20}$ keV~b, respectively.
It should be stressed, however, that the AV18 model, in contrast to the
AV14/UVIII, does not reproduce the experimental binding energies and low-energy
scattering parameters of the three- and four-nucleon systems.
The AV14/UVIII prediction is only 6 \% larger than the AV18/UIX
zero-energy result.  This 6 \% variation should provide
a fairly realistic estimate of the theoretical uncertainty due to
the model dependence.  

\begin{figure}[bth]
\centerline{
\epsfig{file=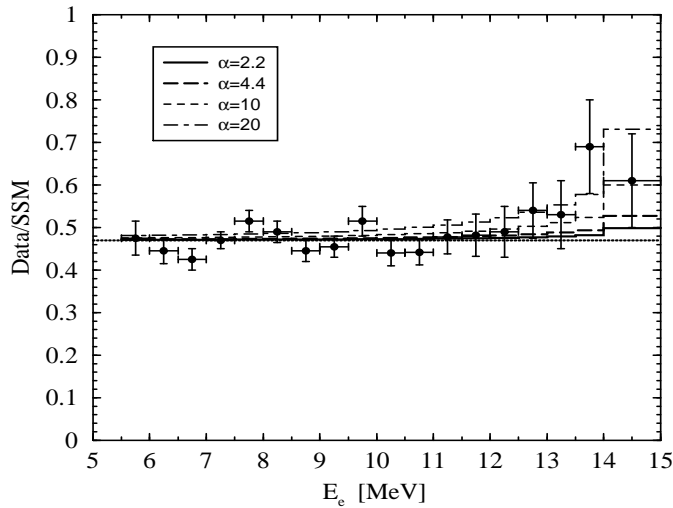,height=2.7in,width=3.5in}}
\caption{Electron energy spectrum for the ratio
between the Super-Kamiokande 825-days data and the expectation
based on unoscillated $^8$B neutrinos~\protect\cite{BBP98}.
See text for an explanation of the symbols.}
\label{fig:ratio}
\end{figure}
The implications of these predictions
for the SuperKamiokande (SK) solar neutrino
spectrum are summarized Fig.~\ref{fig:ratio}.
The SK results are presented as the ratio of the measured electron
spectrum to that expected in the SSM with no neutrino oscillations.
Over most of the spectrum, this ratio is constant at $\simeq 0.5$.  At
the highest energies, however, an excess relative to $0.5 \times$SSM
is seen (though it has diminished in successive data sets).  The SK
825-day data are shown by the points in Fig.~\ref{fig:ratio}
(the error bars denote the combined statistical and systematic error).  
In the figure, the ratio of the $hep$ flux to its value in the SSM
(based on the $hep$ S-factor prediction of Ref.~\cite{Sch92}) is
denoted by $\alpha$, defined as
$\alpha \equiv ( S_{\rm new}/S_{\rm SSM} ) \times P_{\rm osc}$,
where $P_{\rm osc}$ is the $hep$-neutrino suppression constant.
Presently, $\alpha = (10.1\times 10^{-20}{\rm\ keV~b})
/(2.3\times 10^{-20}{\rm\ keV~b}) = 4.4$, if
$hep$ neutrino oscillations are ignored.
The lines in Fig.~\ref{fig:ratio}
indicate the effect of
various values of $\alpha$ on the ratio of the electron spectrum with
both $^8$B and $hep$ to that with only $^8$B (the SSM).
In calculating this ratio, the $^8$B flux in
the numerator has been suppressed by 0.47, the best-fit constant value
for the observed suppression.  If the $hep$ neutrinos are suppressed
by $\simeq 0.5$, then $\alpha = 2.2$.  Two other arbitrary values
of $\alpha$ (10 and 20) are shown for comparison.
It appears that the prediction of Ref.~\cite{Mar00}
is unable to explain the distortion observed in the spectrum
at the highest energies.
\section{Outlook}

Improvements in the modeling of two- and three-nucleon interactions and
nuclear electro-weak currents, and the significant
progress made in the last few years in the description of bound and
continuum wave functions, make it now possible
to perform first-principle calculations of interesting low-energy
reactions involving light nuclei.  While the extension
of the CHH technique to treat
systems with $A \ge 6$ may prove difficult, variational and
Green's function Monte Carlo methods should be able to deal
with them effectively, in particular their
pre-capture continuum states.
Experimentally known electromagnetic and weak transitions
of systems in the mass range $4 \leq A \leq 9$ will provide
powerful constraints on models of nuclear currents. 
Work along these lines is being vigorously pursued.
\section{Acknowledgments}
I wish to thank L.E.\ Marcucci, M.\ Viviani, A.\ Kievsky, and S.\ Rosati
for their many important contributions to the work reported here.
I also like to gratefully acknowledge the support of the U.S.\ Department
of Energy under contract number DE-AC05-84ER40150.  

\end{document}